# Magnetic properties of pressurized CsV3Sb5 calculated by using a hybrid functional

Wenfeng Wu[a,b], Xianlong Wang[*,a,b], Zhi Zeng[*,a,b]

[a] *Key Laboratory of Materials Physics, Institute of Solid State Physics, HFIPS, Chinese Academy of Sciences, Hefei 230031, China*
[b] *University of Science and Technology of China, Hefei 230026, China*

———————

Author to whom all correspondence should be addressed: xlwang@theory.issp.ac.cn and zzeng@theory.issp.ac.cn

## Abstract

Based on the hybrid functional, we find that at 0 GPa, the pristine $CsV_3Sb_5$ has local magnetic moment of 0.85 $\mu_B$/unit cell, which is suppressed at pressure of 2.5 GPa resulting in a spin-crossover. Since the ground sate of $CsV_3Sb_5$ with charge density wave (CDW) distortion is non-magnetic state, the local magnetic moment of pristine $CsV_3Sb_5$ will be suppressed by temperature-induced CDW transition at 94 K. The schematic evolution of magnetic moments as functions of pressure and temperature are presented. At low temperature, $CsV_3Sb_5$ is a rare example of materials hosting pressure-induced local magnetic moment, and we suggeste that the effects of local magnetic moments should be considered for understanding its properties.

# 1. Introduction

The kagome lattice with its unique geometric structure has attracted much attention in recent years and the inherent geometric frustration makes it have strong quantum fluctuations at low temperature, which makes people think that it is one of the most likely systems to realize quantum spin liquids [1,2]. Recently, a new family of kagome metals $AV_3Sb_5$ (A=K, Rb, Cs) has received widespread attentions [3]. It has many interesting physical properties such as charge density waves (CDW) [4–15], superconductivity [3–5,9,16–23], non-trivial topological states [7,14,24–26] and so on. It has also been found that with temperature-decreasing, the ground state of this kagome materials will transfer from pristine pattern to the Inverse Star of David (ISD) configuration with CDW, and the nontrivial topological properties are maintained in the ISD configuration [11]. For example, the CDW transition temperature of $CsV_3Sb_5$ was 94 K and it was found to be a $\mathbb{Z}_2$ topological metal with a superconducting ISD ground state [26]. Interestingly, the superconducting transition temperature ($T_c$) of the pressurized $CsV_3Sb_5$ will exhibit a two-domes-like variation [16,17]. With pressure increasing, two superconductivity phases appear at the pressure range of 0 ~ 10 GPa and 15 ~ 100 GPa, respectively [17]. However, the superconducting mechanism of $AV_3Sb_5$ are still very controversial, and some researchers thinks it is a traditional superconductor [27,28], but others think it is non-traditional superconductor [6,8,11,29,30].

Except the superconducting mechanism, whether $AV_3Sb_5$ has a local magnetic moment is also a controversial issue [3,20,22,31,32], and their superconductivity behaviors may have a close relation with the magnetic properties if the local magnetic moment exist. Some experiments show that these materials have magnetic fluctuations and orbital order [3,22,31], while others do not observe local magnetic moments [20,32]. Interestingly, the first-principles simulations based on the generalized gradient correction approximation (GGA) functional show that non-magnetic (NM) state is the ground state of $CsV_3Sb_5$, however, the state with local magnetic moments become more stable than NM state after including the Hubbard $U$ correction by considering moderate electron-electron correlation effect [27]. Please note that, the density functional theory (DFT) plus dynamical mean-field theory (DMFT) calculations found that $KV_3Sb_5$ family materials are weakly correlated metals [28]. The magnetic moment of $CsV_3Sb_5$ will be suppressed by pressure, and the pressure-induced spin-crossover (SC) was observed, which was believed to be the reason of two-domes-like superconductivity variation [27]. However, the magnetic moment and SC behaviors of $CsV_3Sb_5$ depend strongly on the applied $U$ values. For example, when $U = 0.5$ eV ($U = 2.0$ eV), the magnetic moment is 0.2 $\mu_B$ (0.7 $\mu_B$) at 0 GPa, which will be suppressed at the pressure of ~ 10 GPa (~ 40 GPa) [27]. The reason for this is that $U$ is a semi-empirical parameter and sensitive to the crystal field and spin state [33,34]. Therefore, based on the GGA+$U$ method, it is difficult to answer following questions: How large is the magnetic moment of $CsV_3Sb_5$ at 0 GPa? How does the magnetic moment evolve with pressure?

Our previous works shown that the hybrid functional can be used to accurately describe the SC behaviors of the magnetic materials with strongly correlated interactions by using uniform parameter [35–37], since the exchange-correlation functional is improved via inducing correction in the exchange-term in the hybrid functional [38]. Actually, the hybrid functional has been successfully used to describe the properties of many materials with strongly correlated interactions [39–44]. Therefore, in this work, we used hybrid functional to answer the above two questions. Our results show that at 0 GPa the pristine pattern has a magnetic moment while NM is the ground state of ISD configuration, and the magnetic moment of pristine $CsV_3Sb_5$ will be suppressed by applying 2.5 GPa pressure. The pressure-temperature phase diagram of magnetism is presented.

## 2. Methods

The calculations are carried based on the density functional theory (DFT) implemented in Vienna *ab initio* Simulation Package (VASP) [45,46]. The Perdewburke-Ernzerhof (PBE) functional of the generalized gradient correction approximation (GGA) is used for structural relaxation [47]. The simulations of electronic and magnetic properties are carried based on the Heyd-Scuseria-Ernzerhof functional (HSE06) [48] with the separation parameter of 0.2 and the default mixing parameter of $AEXX$ = 0.25. The cutoff energy for plane wave expansion is set to 350 eV. The convergence thresholds of energy and force are set to $1 \times 10^{-5}$ eV and 0.01 eV·Å$^{-1}$. The brillouin zone integral adopts the Monkhorst-pack method, and the **k**-point grid is set as $8 \times 8 \times 4$ for pristine CsV$_3$Sb$_5$, $3 \times 3 \times 3$ for ISD CsV$_3$Sb$_5$. For both SD and ISD cases, the experimentally reported crystal structures [11,17] are used as the initial structures, and the lattice constants and internal atomic positions are fully relaxed. An initial magnetic moment of 1 $\mu_B$ is set to per vanadium atoms.

## 3. Results and discussion

The crystal structures of pristine CsV$_3$Sb$_5$ and ISD CsV$_3$Sb$_5$ with CDW distortion are shown in Figure 1. Firstly, we investigate the properties of pristine CsV$_3$Sb$_5$ with and without spin-polarization at 0 GPa using the GGA-PBE functional. For the spin-polarized structural relaxations, the calculated lattice constants of a = 5.50 Å and c = 9.89 Å are obtained, which is comparable to previous simulations (a = 5.50 Å and c = 9.87 Å, [26]) and experimental values (a = 5.50 Å and c = 9.89 Å, [17]). The unite cell containing 3 vanadium atoms have a very small magnetic moment of 0.153 $\mu_B$ corresponding to 0.027 $\mu_B$/vanadium atom. The none spin-polarized structural simulation results in comparable lattice constants (a = 5.50 Å and c = 9.83 Å) with that of the spin-polarized structural relaxation. Furthermore, the total energy of none spin-polarized structure is marginally (0.4 meV/unit cell) higher than that of spin-polarized structure, which indicates that based on the GGA functional, NM is the ground state of pristine CsV$_3$Sb$_5$. The results consisting with previous theoretical reports [27,49].

However, the calculations using the hybrid functional show that the spin-polarized pristine CsV$_3$Sb$_5$ with a magnetic moment of 0.85 $\mu_B$/unit cell has lower energy than the counterpart without spin-polarization. The magnetic moment of 0.85 $\mu_B$/unit cell is comparable to the calculated magnetic moment using GGA+$U$ ($U$ = 2 eV) functional [27], indicating that the reasonable $U$ value of pristine CsV$_3$Sb$_5$ at 0 GPa should be ~ 2 eV. Since calculations using the more advanced functionals (hybrid functional and GGA+$U$ [27]) than GGA show that spin-polarized pristine CsV$_3$Sb$_5$ is more stable, we believed that at 0 GPa, pristine CsV$_3$Sb$_5$ should has local magnetic moments.

Furthermore, we will discuss the results related to ISD CsV$_3$Sb$_5$. After fully structural relaxations, the simulations based on the GGA functional show that ISD CsV$_3$Sb$_5$ does not have magnetic moments, and NM is its ground state. Following HSE calculations confirm this result. Based on our HSE results of pristine and ISD cases, we can get following conclusion: The pristine CsV$_3$Sb$_5$ has local magnetic moments at room-temperature and 0 GPa, and it will be suppressed by the CDW induced phase transition from pristine to ISD appeared at the low temperature of 94 K.

In Figure 2, the electronic structures calculated based on the hybrid functional are presented. The partial density of states (PDOS) of pristine CsV$_3$Sb$_5$ and ISD CsV$_3$Sb$_5$ at 0 GPa are shown in Figure 2 (a) and (b), respectively. For the PDOS of pristine CsV$_3$Sb$_5$, the *d*-orbitals of V are the main contributor near the Fermi level while the *p*-orbitals of Sb are the second, and strong hybridization between V-*d* and Sb-*p* orbitals can be found. Cs plays the role of providing electrons in the system, and it almost do not contribute to the density of states near the Fermi level. The total DOS shows that the pristine CsV$_3$Sb$_5$ has spin polarization, and the electrons in V *d*-orbitals of large spin polarization. Furthermore, from the spin charge density shown in Figure 2(c), it is clear that the magnetic moment is predominantly contributed by V atoms. In the case of ISD CsV$_3$Sb$_5$ (Figure 2(b)), the spin-up DOS is identified to that of spin-down resulting in the non-magnetic state.

In order to explore the variation of the magnetic moment of pristine CsV$_3$Sb$_5$ under high-pressure, we calculated the magnetic moment of the system under different pressures, and results are shown in Figure 3. With pressure increasing from 0 GPa to 2 GPa, the magnetic moment decreases from 0.85 $\mu_B$ to 0.64 $\mu_B$. At 2.5 GPa, the magnetic moment is suppressed suddenly to 0.08 $\mu_B$ resulting in a spin-crossover. When the pressure continues to increase to 5 GPa, the magnetic moment is completely suppressed by the pressure. In fact, the system maintains non-magnetic state until 10 GPa. The pressure-induced SC is also observed in the pristine CsV$_3$Sb$_5$ by using the DFT+$U$ method [27], however, the SC pressures are generally larger than 10 GPa and sensitive to the applied *U* values. Up to here, we can conclude that at ambient condition, the pristine $CsV_3Sb_5$ has local magnetic moments, which will be suppressed by temperature or pressure. In other words, the SC of pristine $CsV_3Sb_5$ can be induced by not only increasing pressure but also decreasing temperature.

A schematic *P-T* phase diagram of $CsV_3Sb_5$ is shown in Figure 4, where the red shade represents the magnitude of the magnetic moment, and the white area in the right side represents that the magnetic moment is suppressed by pressure. Since previous experimental works shown that the CDW state will be suppressed by high temperature (94 K at 0 GPa) [26] or high pressure (0.8 GPa at 8.9 K) [15], the phase boundary between ISD and pristine configurations is schematically indicated by an arc connecting the *P-T* conditions of (94 K & 0 GPa) and (0 K & 0.8 GPa). Therefore, the white fan-shaped area in the lower-left corner represents that the magnetic moment of the system is suppressed by the CDW. Note that, our hybrid functional calculations show that similar to the case at 0 GPa, the ISD phase also do not have local magnetic moment at 0.8 GPa.

From Figure 4, we can find that at high temperature region (> 94 K), the system is in the pristine state. As the pressure increases, the system maintains a high spin state until 2.5 GPa, spin-crossover occurs around 2.5 GPa giving rise to a low spin state. At low temperature (< 94 K), the CDW induced non-magnetic ISD configuration is the ground state. The CDW will be suppressed by the pressure at ~ 0.8 GPa, and the local magnetic moments appear in the pressure-induced pristine $CsV_3Sb_5$. Actually, whether pressure can induce magnetism is a very interesting question in the field of high-pressure research. At low temperature, $CsV_3Sb_5$ is a rare example acquiring pressure induced local magnetic moments. Furthermore, with pressure increasing to 2.5 GPa, the magnetic moment is suppressed by pressure. In fact, our findings still have some issues to be clarified. For example, at low temperatures, why does CDW inhibit the magnetic properties of the system? How does magnetism affect superconductivity? These problems urgently need to be further resolved. Simultaneously, our work is waiting for experimental verification.

## 4. Summary

In summary, we use the HSE06 functional to study the magnetic properties of pristine $CsV_3Sb_5$ and ISD $CsV_3Sb_5$ structures, and the evolution of magnetic moment with pressure is illustrated. Our results show that at 0 GPa, the pristine $CsV_3Sb_5$ has local magnetic moment, which is suppressed by applying 2.5 GPa pressure resulting in a spin crossover. The ISD $CsV_3Sb_5$ does not have local magnetic moment, indicating that the local magnetic moment of pristine $CsV_3Sb_5$ will be suppressed by temperature-induced CDW transition at 94 K combined with the phase-transition from pristine to ISD configurations. Furthermore, at low-temperature, the magnetic moment of pressurized $CsV_3Sb_5$ can be triggered by the phase transition from ISD to pristine pattern combined with the suppression of CDW by pressure, indicating that $CsV_3Sb_5$ is a rare example of materials hosting pressure-induced magnetism. To understand the properties of $CsV_3Sb_5$ sufficiently, the effects of local magnetic moment should be considered.

## Acknowledgements

This research was supported by the NSFC under Grant of U2030114. Science Challenge Project No. TZ2016001. The calculations were partly performed in Center for Computational Science of CASHIPS, the ScGrid of Supercomputing Center and Computer Network Information Center of Chinese Academy of Sciences, and partly using Hefei advanced computing center.

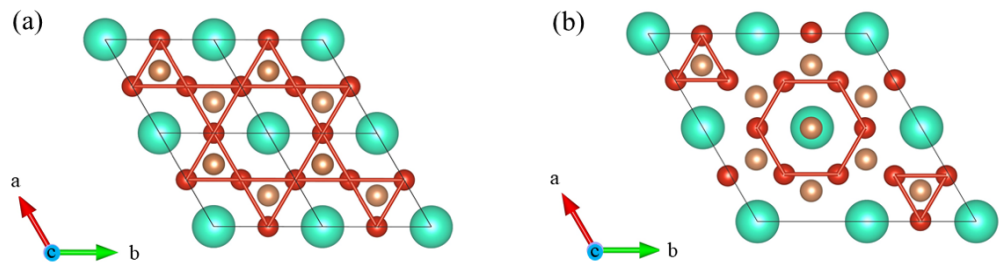

**Figure 1.** (a) Structure of pristine $CsV_3Sb_5$. (b) Structure of $CsV_3Sb_5$ with charge density wave (CDW) distortion (2x2 ISD pattern). The green, brown and red spheres denote Cs, Sb, V atoms, respectively.

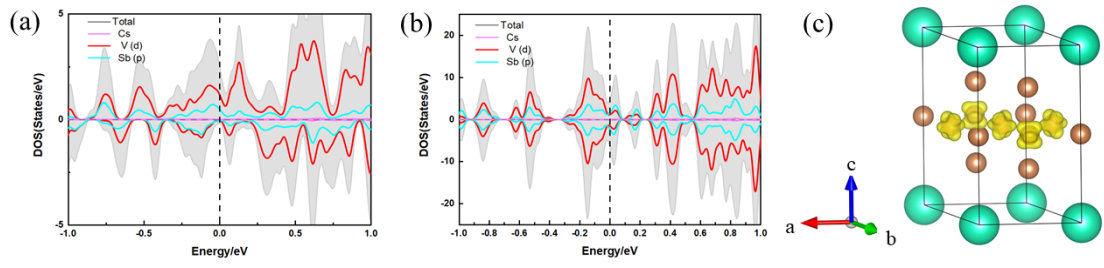

**Figure 2.** (a) and (b) show the DOS of pristine $CsV_3Sb_5$ and ISD $CsV_3Sb_5$. (c) Spin charge density of pristine $CsV_3Sb_5$, iso-surface at 0.01 e/Å$^3$.

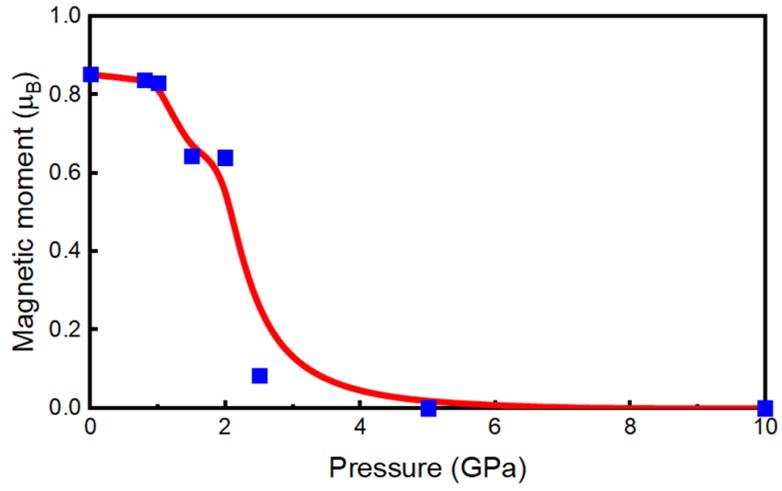

**Figure 3.** The magnetic moments of pristine CsV$_3$Sb$_5$ are shown as a function of pressure, which are obtained based on the hybrid functional.

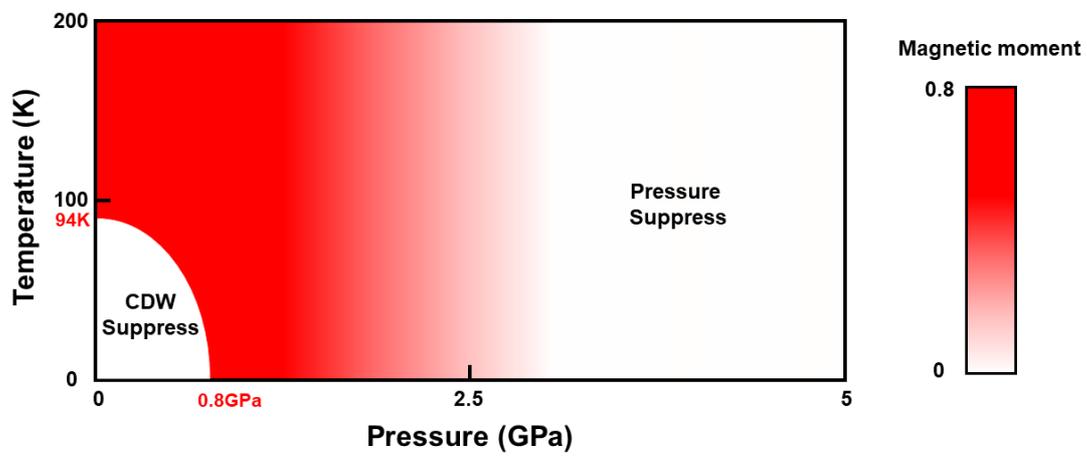

**Figure 4.** Schematic *P-T* phase diagram of CsV$_3$Sb$_5$. The intensity of red represents the magnitude of the magnetic moment, and the gray area represents the suppression of magnetism by CDW. The white area represents magnetism is suppressed by pressure.